 \documentclass[preprint,aps]{revtex4}
 
\begin{document}
 \newcommand{\Qed}{\rule{2.5mm}{3mm}}
 \newcommand{\balpha}{\mbox{\boldmath {$\alpha$}}}
 \draft 
%\documentstyle[preprint,aps]{revtex}
%\begin{document}
%\draft
\title{Mass Protection and No Fundamental Charges Calls for Extra Dimensions}
\author{ N.S. Manko\v c Bor\v stnik}
\address{ Department of Physics, University of
Ljubljana, Jadranska 19,Ljubljana, 1111, \\
and Primorska Institute for Natural Sciences and Technology,\\
C. Mare\v zganskega upora 2, Koper 6000, Slovenia}
\author{ H. B. Nielsen}
\address{Department of Physics, Niels Bohr Institute,
Blegdamsvej 17,\\
Copenhagen, DK-2100
 }

%\date{\today}

%\maketitle

\begin{abstract} 
We call attention to that assuming %that there are 
no  conserved
charges in the fundamental theory, but rather only gravity and fermions 
with only a spin,
the dimensions 4, 12, 20, ... 
are excluded under the requirement of 
mass protection. If it is required that we shall have several families 
the generic result is that even the other by 4 devisable dimensions are excluded
and indeed only d=2 (mod 4) remains as acceptable.  
\end{abstract}

\maketitle

%\newpage
\date{today}

\section{Introduction}
\label{introduction}

Looking at the Standard model of the electroweak and colour interaction,
it is well known that it is strongly suggested 
that all the particles in this model - except for the Higgs particle itself -
are a priori massless and only obtain masses different from zero by means
of the interaction with the Higgs field (its vacuum expectation value). 
Especially the fermions - quarks
and leptons - are in this sense mass protected, because their fields are 
composed from (a couple of) Weyl fields with such quantum numbers under the 
``weak'' gauge transformations that a mass term is forbidden unless the 
interaction with the Higgs field vacuum expectation value causes the mass term, 
that means that the Higgs field "dresses" the right handed weak chargeless fermions 
with a weak charge.

The main point to which we call attention here is, however, that this 
mass protection {\em only works when there exist in the theory a} (for instance
gauge) {\em symmetry ensuring the conservation of one or several  charges 
distinguishing the right and the left handed Weyl-components.} Here, however, 
we want to look for, how it would go if we assume that there exists no such 
charge conservation a priori, or let us say at the fundamental level.
Indeed if the charge conservation were broken, then there would be the 
allowance/possibility for a mass term being added to the 
Lagrange density. Even for the left handed neutrino, which in the Standard
model does not have any right handed partner making it able to obtain
a Dirac mass term, there is the possibility of a Majorana term provided 
the conservation laws are broken.

Really the possibility of Majorana terms is so general that we quite 
generally can claim that there is no way to get mass protection for 
any (Weyl) fermion without use of the charge conservation in the usual 
3+1 dimensions. However, this statement is dimension-dependent, and the point 
is that we shall see that the dimensions which modulo 8 are equivalent 
to 4 are excluded under the assumption of there being no charge conservation
imposed. Actually we shall see that it is even so that the potential Majorana
mass terms for Weyl particles in dimensions divisible by 8 is only excluded 
even for no charge conservation because of the potential term having to be 
symmetric under the permutation of the (second quantized) fermion fields. 
Because of the Fermi statistics such a symmetry is not allowed and thus the
mass protection at first looks to be possible, but this is only  true as
long as  there is only one family/flavour. When there are more families of
Weyl particles however even in by 8 divisible dimensions only one of these
families will survive as massless. So if we insist on the number of families
being mass protected should be bigger than one even these by 8 divisible 
dimensions are excluded under the assumption of no conservation of charges.

Since it is  well known  
and rather trivial that there cannot be mass protection in an odd number 
of space-time dimensions\cite{hn00,hn02},  since in odd dimensions Weyl spinors are at 
the same time Dirac spinors, we are at the end left with that under the 
assumption of there being no conserved charges we can only have mass protection
of more than one family for the dimensions d = 2 (mod 4); i.e. for such 
dimension numbers as 2,6,10,14,...

One can of course complain against this discussion by saying  that it is 
quite contrary to what we know to assume that there are no conserved charges:
One of us have long worked on the speculation that the conservation 
laws which we conceive of as charges phenomenologically are at a deeper level
to be identified with angular momenta in the extra dimensions\cite{norma92,norma93,norma95,Portoroz03} 
(more precisely with spin degrees of freedom). 
Taken the generic point of view of not taking there to be a charge conservation
without reason we see that one is first driven to some dimension having 
2 as remainder when divided by four and then in order to obtain the 
phenomenologically known charges a scheme like the one mentioned of obtaining 
them as angular momenta (spin) would be highly suggested.

\section{Massless and massive fermions}
\label{masses}

Let $d$ be any integer number. We pay attention to only spinors (fermions). 
We see that: \\
i) In $d$ even the operator of 
handedness $\Gamma^{d}$ ($(\Gamma^{d})^2=I,\; \Gamma^{d \dagger}= \Gamma^{d}$) 
is a Clifford even operator, proportional to an even number 
of all the Clifford odd 
objects $\gamma^a$ and can accordingly be expressed as a product of 
all the $d/2$ elements of the Cartan subalgebra set of the Lorentz generators $S^{ab}$ for spinors.
One finds
\begin{eqnarray}
\label{gammacomdeven}
\{\Gamma^{d}, \gamma^a\}_+ =0, \quad \{\Gamma^{d}, S^{ab}\}_- =0,\quad {\rm d \; even}.
\end{eqnarray}
ii) For  $d$ an  odd number the handedness is a Clifford odd operator, proportional 
to a product of an odd number of $\gamma^a$, which means that it can be chosen to be 
proportional to the product of all the $(d-1)/2$ members of the Cartan subalgebra set 
of the commuting generators of the Lorentz group and one of $\gamma^a$, say  
$\gamma^d$. This means that $\Gamma^{d}$ changes 
the Clifford character of a state, if being applied on a state of a definite Clifford 
character. One finds
\begin{eqnarray}
\label{gammacomdodd}
\{\Gamma^{d}, \gamma^a\}_- =0, \quad \{\Gamma^{d}, S^{ab}\}_- =0,\quad {\rm d \; odd}.
\end{eqnarray}
The mass term has in a second quantized formalism a form
\begin{eqnarray}
\label{mass0}
-m {\bf \hat{\Psi}}^{\dagger} \gamma^0 {\bf \hat{\Psi}},
\end{eqnarray}
where ${\bf \hat{\Psi}}$ is the operator annihilating a fermionic state
\begin{eqnarray}
\label{spinorstate}
<0|{\bf \hat{\Psi}}(x)| \Psi_k> = \Psi_k(x).
\end{eqnarray}
\subsection{Spinors with no charge and no family}
\label{nofamily}

{\em We assume that a spinor  carry nothing but a spin and that  only 
a Weyl of one handedness and  one family index exists.}

{\em Statement 1}: There is no Dirac mass term for $d$ even and it is always a Dirac 
mass term for $d$ odd. 

{\em Proof:}
The proof is straightforward and well known. Namely, choosing only one handedness, say left, 
the mass term has a form $-m ((1-\Gamma^{d}){\bf \hat{\Psi}})^{\dagger} \gamma^0 
((1-\Gamma^{d})){\bf \hat{\Psi}} =-m {\bf \hat{\Psi}}^{\dagger} (1-\Gamma^{d})\gamma^0 (1-\Gamma^{d}) 
{\bf \hat{\Psi}}$, which is zero (because of the factor $\gamma^0(1+\Gamma^{d})(1-\Gamma^{d})$) 
for $d$ even and nonzero (the factor is now $\gamma^0 (1-\Gamma^{d})$) for $d$ odd, due to Eqs.
(\ref{gammacomdeven},\ref{gammacomdodd}).

{\em Statement 2}: There is no Majorana mass term for 
$d =2(2k+1)$, $k=0,1,2,\cdots $ and is also no Majorana mass term for $d =8k$, $k=1,2,\cdots,$ for 
only one family of spinors.  
In all other dimensions there is always the Majorana mass term.

We shall prove this statement in several steps.

Let ${\bf \hat{{\cal C}}}$ be a charge conjugation operator, operating on ${\bf \hat{\Psi}}$. 
Then the creation operator creating a Majorana state out of any left handed state, 
is defined as follows 
\begin{eqnarray}
\label{majorana}
{\bf \hat{\Psi}}_{M} = \frac{1}{\sqrt{2}}({\bf \hat{{\cal C}}}(1- 
\Gamma^d){\bf \hat{\Psi}}
{\bf \hat{{\cal C}}}^{-1} \pm (1- \Gamma^d){\bf \hat{\Psi}}).
\end{eqnarray}
Let us write $d= 4n +2m$ for $d$ even and $d= 4n +2m -1$ for $d$ odd, with $n=1,2,\cdots$, 
and with $m=0,1$ only.

{\em Statement 2a}: The following relation holds
\begin{eqnarray}
\label{majoranarelation}
{\bf \hat{{\cal C}}}(1-\Gamma^d){\bf \hat{\Psi}}{\bf \hat{{\cal C}}}^{-1} =
(-1)^{n-1+m} \prod_{Im \;\gamma^a} \gamma^a \;((1-\Gamma^d) {\bf \hat{\Psi}})^{\dagger}.
\end{eqnarray}
(The index $Im\; \gamma^a$ under the product means that one takes the product 
over those $\gamma^a $-martices which have only imaginary matrix elements.)
This statement is proven in the subsect. \ref{statement2a}.

We make a choice of $\gamma^a$ so that the first two ($\gamma^0, \gamma^1$) are real, $\gamma^2$ 
is imaginary, $\gamma^3$ is real, (we skip index $4$) $\gamma^5$ is imaginary and 
all the rest  $\gamma^a$ with $a$ 
even are real and those with $a$ odd are imaginary. We have $(\gamma^a)^{\dagger} = \eta^{aa} \gamma^a$ 
and $\eta^{aa}= diag(1,-1,-1,\cdots\;)$. (Accordingly $\gamma^0$ is a symmetric $2^{(d/2-1)} 
\times 2^{(d/2-1)} $ matrix and so are all the imaginary 
$\gamma^a$'s, while the real $\gamma^a$ matrices (except $\gamma^0$) are antisymmetric.) 
One finds that the number of imaginary $\gamma^a$ in the product
$\prod_{Im \gamma^a} \gamma^a $ is for $d$ even equal to $\frac{d-2}{2}$ and for $d$ odd 
$\frac{d-1}{2}$.
To see in which dimensions  the Majorana mass term 
\begin{eqnarray}
\label{majoranamass0}
-m {\bf \hat{\Psi}}^{\dagger}_{M} \gamma^0 {\bf \hat{\Psi}}_M,
\end{eqnarray}
gives zero, if spinors of only one handedness and one family are assumed, we must evaluate  the part 
\begin{eqnarray}
\label{majoranarelation1}
-(\pm)\; m\{({\bf \hat{{\cal C}}}(1-\Gamma^d){\bf \hat{\Psi}}{\bf \hat{{\cal C}}}^{-1})^{\dagger} \gamma^0
\;((1-\Gamma^d) {\bf \hat{\Psi}}) + ((1-\Gamma^d) {\bf \hat{\Psi}})^{\dagger} \;
({\bf \hat{{\cal C}}}(1-\Gamma^d){\bf \hat{\Psi}}{\bf \hat{{\cal C}}}^{-1})\}= \nonumber\\
-(\pm)\; m\{{\bf \hat{\Psi}}(1-\Gamma^a)(\prod_{Im \gamma^a} \gamma^a )^{\dagger} \gamma^0
\;(1-\Gamma^d) {\bf \hat{\Psi}} + ({\bf \hat{\Psi}})^{\dagger} (1-\Gamma^d) \gamma^0 
\;(\prod_{Im \gamma^a} \gamma^a) (1-\Gamma^d){\bf \hat{\Psi}}^{\dagger}\}. 
\end{eqnarray}
It is easy to see that the term of Eq.(\ref{majoranarelation1}) is zero for $d = 2(2k +1)$, for 
$k=0,1,2,\cdots\;$, since $(1-\Gamma^d)\; (\prod_{Im \gamma^a} \gamma^a)^{\dagger} \gamma^0 
(1-\Gamma^d)= (\prod_{Im \gamma^a} \gamma^a)^{\dagger} \gamma^0 (1 + \Gamma^d)
(1-\Gamma^d) =0$ and similarly also the term $(1-\Gamma^d)\; \gamma^0 (\prod_{Im \gamma^a} \gamma^a)  
(1-\Gamma^d)=0,$ 
namely in even dimensional spaces $\Gamma^d$ anticommutes with a product of an odd number of 
$\gamma^a$'s (and it commutes with a product of an even number of $\gamma^a$'s). 
%, while in odd dimensional spaces $\Gamma^d$  commutes with either an 
%even or an odd number of $\gamma^a$'s. 
For $d=8k$, however, we still get a zero contribution, 
due to the fact that the two terms: the term  
$(\prod_{Im \gamma^a} \gamma^a )^{\dagger} \gamma^0
\;(1-\Gamma^d)$ and his Hermitean conjugate one $ \gamma^0 \prod_{Im \gamma^a} \gamma^a 
\;(1-\Gamma^d)$ cancel each other. 
%This will however not happen if there were more than one family.

We can conclude: {\em In $d=2(mod\;  4)$ and $d=0(mod\;8)$ dimensions, if we only have a 
Weyl of one handedness and 
one family and no charges,   
there is  no Dirac mass term and also  no Majorana mass term.}

In all the odd dimensions $d$ the Dirac term by itself gives a nonzero contribution what ever the 
Majorana term is. But since in odd dimensional spaces $\Gamma^d$ commutes with any product of 
$\gamma^a$, also Eq.(\ref{majoranamass0}) gives a nonzero contribution.

%We must check whether terms of Eq.(\ref{majoranarelation1}) can still give a zero contribution 
%because of some
%symmetry reasons. And indeed we find that the Majorana mass term is nonzero in $d=0 (mod  4)$, while it 
%gives a zero contribution in  $d=0 (mod  8)$, but only if there is no additional families and no charges. 

%
\subsection{Spinors with no charge appearing in families}
\label{family}

If we allow a family index without any special requirement about  global or local symmetries of 
our Lagrangean with respect to the family index, it will in general happen that in 
$d=0(mod\;8)$ dimensions the two terms of Eq.(\ref{majoranarelation1}) do not cancel each other, 
so that the mass term of the Majorana type of Eq.(\ref{majorana}) will be non zero. 

Really one should imagine that we have a Majorana like  mass term given by a matrix in 
the space of families $m^{M}_{fg}$ of the form
\begin{eqnarray}
\label{familymajoranamass0}
-m^{M}_{fg} {\bf \hat{\Psi}}^{\dagger}_{M f} \gamma^0 {\bf \hat{\Psi}}_{M g},
\end{eqnarray}
and it is seen that the cancellation takes place for the symmetric part of the 
family matrix $m^{M}_{fg}$. This means that if the number of families is {\em odd}
there has to be a zero-eigenvalue for the antisymmetric part of  this matrix 
and thus a single family will survive to have zero mass. 

We can conclude: {\em It is only  $d=2(mod\;  4)$  dimensional spaces that Weyl 
spinors of one handedness,  no charge and more than one family index are mass protected, 
having no  Dirac mass term and also  no Majorana mass term.}

\subsection{The proof for the statement 2a}
\label{statement2a}

\noindent
i. Let $X^{p}_> : = \{\Phi_{k >}\}$ be a set of $p$ occupied single particle states 
above the Dirac sea and let 
 $X^{p}_<: = \{\Phi_{l <}\}$ be a set of $r$ holes in  the Dirac sea. \\
ii. Let us make 
a choice of any phase for $\Phi_{k >}$, while we choose phases for $\Phi_{l <}$ so that
\begin{eqnarray}
\label{phase}
C \Phi_{l >} = \Phi_{l <}, \quad
C = (\prod_{Im \gamma^a}\; \gamma^a) \; K,
\end{eqnarray}
with an odd number of $\gamma^a$'s in $d= 0 (mod \;4)$ and $d=0 (mod \;4)-1$ and with  
an even number of $\gamma^a$'s in $d= 2 (mod \;4)$ and $d= 0 (mod \;4)-1$. K is an antilinear operator, 
transforming a complex number into its complex conjugate number.
Then
\begin{eqnarray}
C^2 \Phi_{k >} &=& C \Phi_{n <} = (-)^{n-1 +m} \Phi_{n >},\quad  \nonumber\\
{\rm with } \;d &=& 4n+2m -1, \;{\rm for\;} d \; {\rm odd},\nonumber\\
{\rm and}\; d&=& 4n+2m, \;{\rm for\;} d \; {\rm even},\;
m=0,1; n=0,1,2,\cdots\;.
\label{C}
\end{eqnarray}
We find $X^{p+1}_>  = X^{p}_> U  \{\Phi_{k >}\}= \Phi_{1 >},\Phi_{2 >}, \cdots, \Phi_{k >},
\Phi_{p >} . (-)^{a_k X^{p}_{>}}$, with $a_k X^{p}_{>}=0,$ if $\Phi_{l >}$ jumps over an even number 
of $\Phi_{i >}$ and $a_k X^{p}_{>}=1,$ if $\Phi_{k >}$ jumps over an odd number 
of $\Phi_{i >}$. $X^{p}_> U  \{\Phi_{k >}\}= 0$ for $  \{\Phi_{k >}\} \in X^{p}_> $ and 
equivalently for $X^{r}_< U  \{\Phi_{r <}\}$. 
We further find 
\begin{eqnarray}
{\bf \hat{{\cal C}}} |X^{p}_>, X^{r}_< > = 
(-)^{(m+n-1)r} (-)^{p r} | X^{r}_<, X^{p}_> >.
\label{Candvacuum}
\end{eqnarray}
We define the creation operator for a spinor (fermion) state $\Phi_{k >}$ above the Dirac sea as follows
\begin{eqnarray}
\label{creation}
{\bf \hat{b}^{\dagger}}_{\Phi_{k >}} |X^{p}_>, X^{r}_< > &=& |X^{p}_> U \{\Phi_{k >}\}, X^{r}_< > 
= (-)^{a_k X^{p}_>} |X^{p+1}_>, X^{r}_< >,\; 
{\rm if}\; \Phi_{k >} \notin X^{p}_>,\nonumber\\
{\bf \hat{b}^{\dagger}}_{\Phi_{k >}} |X^{p}_>, X^{r}_< > &=& 0,\;  {\rm if} \;\Phi_{k >}
\in X^{p}_>. 
\end{eqnarray}
We define the creation operator for a hole state, that is the annihilation operator 
which annihilates the state $\Phi_{l <}$ in the Dirac sea as follows
\begin{eqnarray}
\label{annihilation}
{\bf \hat{b}}_{\Phi_{l <}} |X^{p}_>, X^{r}_< > &=& (-)^p |X^{p}_> , X^{r}_< > U \{\Phi_{l <}\} = 
(-)^{a_{l X^{r}_<} +p } |X^{p}_>, X^{r+1}_< >,\; {\rm if}\; 
\Phi_{l <} \notin X^{r}_<,\nonumber\\
{\bf \hat{b}}_{\Phi_{l <}} |X^{p}_>, X^{r}_< > &=& 0,\;  {\rm if} \;\Phi_{l <}
\in X^{r}_<. 
\end{eqnarray}
Equivalently we define the annihilation operator annihilating a  spinor state $\Phi_{k >}$ 
above the Dirac sea as 
\begin{eqnarray}
\label{annihilationabove}
{\bf \hat{b}}_{\Phi_{k >}} |X^{p}_>, X^{r}_< > &=&  
 (-)^{a_k X^{p}_< } |X^{p-1}_>, X^{r}_< >,\; 
{\rm if}\; \Phi_{k >} \in X^{p}_>,\nonumber\\
{\bf \hat{b}}_{\Phi_{k >}} |X^{p}_>, X^{r}_< > &=& 0,\;  {\rm if} \;\Phi_{k >}
\notin X^{p}_>,  
\end{eqnarray}
while we define the annihilation operator annihilating a hole in the Dirac sea by 
\begin{eqnarray}
\label{annihilationbellow}
{\bf \hat{b}^{\dagger}}_{\Phi_{l <}} |X^{p}_>, X^{r}_< > &=&  
 (-)^{a_l X^{r}_< +p} |X^{p}_>, X^{r-1}_< >,\; 
{\rm if}\; \Phi_{k >} \in X^{r}_>,\nonumber\\
{\bf \hat{b}^{\dagger}}_{\Phi_{l <}} |X^{p}_>, X^{r}_< > &=& 0,\;  {\rm if} \;\Phi_{l <}
\notin X^{r}_<. 
\end{eqnarray}

\noindent
{\em Statement 2 b 1:} 
\begin{eqnarray}
\label{CalConb++}
{\bf \hat{{\cal C}}} \;{\bf \hat{b}}^{\dagger}_{\Phi_{k >}} &=& 
{\bf \hat{b}}_{\Phi_{k <}} \;{\bf \hat{{\cal C}}}. 
\end{eqnarray}
{\em Proof:} Let $X^{p}_>$  not include the state $\Phi_{k >}$. One then 
finds that ${\bf \hat{{\cal C}}}\;{\bf \hat{b}}^{\dagger}_{\Phi_{k >}}
|X^{p}_>,X^{r}_<> = $ $
{\bf \hat{{\cal C}}}$ $(-)^{a_{k}X^{p}_{>}}$ $|X^{p+1}_>,X^{r}_<> = $ $
(-)^{(n-1+m)r}$ $(-)^{a_{k}X^{p}_{>}}$ $(-)^{r(p+1)}|X^{r}_<,X^{p+1}_>>$ $ = $ $(-)^{(n-1+m)r}%
(-)^{a_{k}X^{p}_{>}} $ $(-)^{r(p+1)} (-)^{-a_{k}X^{p}_{>}} $ $(-)^r \;{\bf \hat{b}}_{\Phi_{k <}}$ $
|X^{r}_<,X^{p}_>> = $ $\;{\bf \hat{b}}_{\Phi_{k <}}$ ${\bf \hat{{\cal C}}}$ $
|X^{p}_>,X^{r}_<>$, which completes the proof, since if $X^{p}_>$ does include the 
state $\Phi_{k >}$, the very left hand side  before the first equality sign is zero and so is 
evidently zero on the right hand side one equality sign before the last.

\noindent
{\em Statement 2 b 2:} 
\begin{eqnarray}
\label{CalConb+-}
{\bf \hat{{\cal C}}} \;{\bf \hat{b}}^{\dagger}_{\Phi_{k <}} &=& (-)^{-(n-1+m)}{\bf \hat{b}}_{\Phi_{k >}} 
\;{\bf \hat{{\cal C}}}. 
\end{eqnarray}
{\em Proof:} Let $X^{r}_<$ includes the state $\Phi_{k <}$. One then finds that 
${\bf \hat{{\cal C}}}\;{\bf \hat{b}}^{\dagger}_{\Phi_{k <}}|X^{p}_>, X^{r}_<> = $ $
{\bf \hat{{\cal C}}}$ $(-)^{a_{k}X^{r}_{<}+p}$ $|X^{p}_>,X^{r-1}_< > = $ $
(-)^{(n-1+m)(r-1)} $ $(-)^{a_{k} X^{r}_{<}+p}$ $(-)^{(r-1)p}|X^{r-1}_<,% 
X^{p}_>>$ $ = $ $(-)^{(n-1+m)(r-1)}\;%
(-)^{a_{k}X^{r}_{<}+p}$ $(-)^{(r-1)p}\;(-)^{-a_{k}X^{r-1}_{<}}$ $\;{\bf \hat{b}}_{\Phi_{k >}}$ $
|X^{r}_<, X^{p}_>> = $ $(-)^{n-1+m} \;{\bf \hat{b}}_{\Phi_{k >}}$ $\;{\bf \hat{{\cal C}}}$ $%
|X^{p}_>, X^{r}_<>$, which completes the proof.

\noindent
{\em Statement 2 b 3:} 
\begin{eqnarray}
\label{CalConb+}
{\bf \hat{{\cal C}}} \;{\bf \hat{b}}_{\Phi_{k >}} &=& {\bf \hat{b}}^{\dagger}_{\Phi_{k <}} 
\;{\bf \hat{{\cal C}}}. 
\end{eqnarray}
{\em Proof:} The proof goes equivalently as in the case of Statements 2 b 1 and 2 b 2.

\noindent
{\em Statement 2 b 4:} 
\begin{eqnarray}
\label{CalConb-}
{\bf \hat{{\cal C}}} \;{\bf \hat{b}}_{\Phi_{k <}} &=& (-)^{(n-1+m)}{\bf \hat{b}}^{\dagger}_{\Phi_{k >}} 
\;{\bf \hat{{\cal C}}}. 
\end{eqnarray}
{\em Proof:} The proof goes equivalently as in the case of Statements 2 b 1 and 2 b 2.
 
One finds by repeating the operation with ${\bf \hat{{\cal C}}} $ two times that  
\begin{eqnarray}
\label{CalC2}
{\bf \hat{{\cal C}}}^2 \;{\bf \hat{b}}^{\dagger}_{\Phi_{k >,<}} &=& (-)^{(n-1+m)}
{\bf \hat{b}}^{\dagger}_{\Phi_{k >,<}} 
\;{\bf \hat{{\cal C}}}^2. 
\end{eqnarray}
And we have for the vacuum state
\begin{eqnarray}
\label{CalC2vac}
{\bf \hat{{\cal C}}}^2 \;|X^{p=0}_> , 
X^{r=0}_< >  =  |X^{p=0}_> , 
X^{r=0}_< >. 
\end{eqnarray}

 We can now write for the operator ${\bf \hat{\Psi}}(x)$ the relation 
 \begin{eqnarray}
 \label{spinoroperator}
 {\bf \hat{{\Psi}}}(x) = \sum_{k_<} \; \Psi_{k<}(x) \;{\bf \hat{b}}_{\Phi_{k<}} + 
 \sum_{k_>} \;\Psi_{k>}(x) \;{\bf \hat{b}}_{\Phi_{k >}},\nonumber\\
 {\bf \hat{{\Psi}}}^{\dagger}(x) = \sum_{k_<} \;\Psi^{\dagger}_{k<}(x) \;
 {\bf \hat{b}}^{\dagger}_{\Phi_{k<}} + 
 \sum_{k_>} \;\Psi_{k>}^{\dagger}(x) \;{\bf \hat{b}}^{\dagger}_{\Phi_{k >}}, 
 \end{eqnarray}
with $\Psi^{\dagger}_{k<,k>}(x) $ $= \Psi^{*}_{k<,k>}(x)=$ $ K \Psi_{k<,k>}(x)$, 
that is complex conjugation. 

We further find that (using the relation $C\Psi_{k>} $ $= \Psi_{k<}$ and $C\Psi_{k<} = (-)^{n-1+m}\;$ $
\Psi_{k>}$ (Eq.(\ref{C})) ) ${\bf \hat{{\cal C}}}\;$ ${\bf \hat{{\Psi}}}(x) = $ $\sum_{k<} 
\;\Psi_{k<} (x) $ $\;{\bf \hat{b}}^{\dagger}_{\Phi_{k>}} $ $\; (-)^{n-1+m}$ ${\bf \hat{{\cal C}}}\; + $ $
 \sum_{k_>} \;$ $\Psi_{k>}(x) $ $\;{\bf \hat{b}}^{\dagger}_{\Phi_{k<}}$ $\;{\bf \hat{{\cal C}}}\; $ $=
(-)^{n-1+m} $ $\;  [ \sum_{k<}$ $ 
\; (C \Psi_{k>}(x))$ $ \;{\bf \hat{b}}^{\dagger}_{\Phi_{k > }} \; $ $ + 
 \sum_{k_>} $ $\;(C \Psi_{k<}(x)) $ $\;{\bf \hat{b}}^{\dagger}_{\Phi_{k<}} ]$ $\; {\bf \hat{{\cal C}}}$
 so that it follows 
  \begin{eqnarray}
  \label{spinoroperatorrel}
  {\bf \hat{{\cal C}}}\;{\bf \hat{{\Psi}}}(x) = (-)^{n-1+m} \;   \sum_{{\rm all\;} k} 
\; (C \Psi_{k}(x)) \;{\bf \hat{b}}^{\dagger}_{\Phi_{k}} \;{\bf \hat{{\cal C}}},\nonumber\\
. 
  \end{eqnarray}
Since (Eq.(\ref{C})) $C \Psi (x)$ $= \prod_{IM \gamma^a} \; \gamma^a $ $ \; K \Psi(x)$, we have  
$K \Psi_{k<,k>}(x)$ $= (\prod_{ Im \gamma^a} $ $\; \gamma^a)^{-1}\; $ $C \Psi_{k<,k>}(x)$ and accordingly
  \begin{eqnarray}
  \label{spinoroperatorrel1}
  {\bf \hat{{\cal C}}}\;{\bf \hat{{\Psi}}}(x) = (-)^{n-1+m} \; (\prod_{ Im \gamma^a}\; \gamma^a)
  \; {\bf \hat{{\Psi}}}^{\dagger}(x) \;{\bf \hat{{\cal C}}}, 
  \end{eqnarray}
from where the relation 
\begin{eqnarray}
\label{majoranarelationproof}
{\bf \hat{{\cal C}}}{\bf \hat{\Psi}}{\bf \hat{{\cal C}}}^{-1} =
(-1)^{n-1+m} \; (\prod_{Im \gamma^a} \gamma^a) \; {\bf \hat{\Psi}}^{\dagger}
\end{eqnarray}
follows, proving the Statement 2a of Eq.(\ref{majoranarelation}), since the 
Eq.(\ref{majoranarelationproof}) holds also for $(1-\Gamma^d){\bf \hat{\Psi}}$, 
accordingly relating it with $((1-\Gamma^d){\bf \hat{\Psi}})^{\dagger}$.

\section{Conclusions}       
\label{discussions}    

We have pointed out in this paper that the mass protection mechanism - put into the Standard 
model of the electroweak and colour interaction ''by hand'' with the assumption that only the left 
handed spinors carry the weak charge while the right handed spinors are weak chargeless - 
 {\em only works in $d$ is $(1+3)-$dimensional space, when there exist in the theory a} 
 {\em symmetry ensuring the conservation of one or several  charges 
distinguishing the right and the left handed Weyl-components.} If a spinor of only one handedness 
carries no charge and 
no more than one family index, then the mass protection mechanism works only for $d = 2(2n + 1)$ and 
$d=8n,$ for any $n$. If a spinor of only one handedness and no charge has more than one family 
then the mass protection mechanism works only in dimensions $d = 2(2n + 1)$. Otherwise one gets 
in a generic case only one massless family and that even only in the odd number of families case.

To appreciate this observation as a significant one, one should think in the philosophy put forward 
through many years by one us\cite{norma92,norma93,norma95,Portoroz03}: 
It would be a nice simplification of the theory if instead of many 
gauge fields put "by hand" into the model explicitely only the gravity would exist. 

If one now in this philosophy nevertheless asks for having mass protection so that one can get 
particles with a zero or low mass  to be observed by the 
physicists that have compared to say the Planck scale 
 only very low energy accellerators at their disposal, then he has to ask for how  can one 
get mass protected fermions - let us say  a Weyl spinor - without any conserved 
charge which  assigns differently to right and left components. This was the question we really 
addressed above, and the result turned out to be that {\em one  would need a different 
dimension from the experimental $1+3$ one, namely if one wants to end up with more than 
one family (as we need phenomenologically) a space time dimension} must be $d=2(2n + 1)$.

{\em In other words taking mass protection and no fundamental conserved charges roughly speaking 
as the guiding principles one is driven to fundamentally there being a number of dimensions 
which is equivalent to 2 modulo 4.}

Of course it can very well be that in the rough sense needed here the gauge symmetries 
of the Standard model are fundamantal - and not what is here really what 
not fundamental in this rough sense means, Kaluza-Klein, - since so it is e.g. in 
string theories or  it could appear in many other ways. %\cite{gaugetheoryderivation}.
However, in a sense it complicates the model if we anyway have to have gravity in the theory
and gravity could via the Kaluza Klein mechanism produce the gauge fields, to then put 
in explicite gauge fields. 

Now it must however be admitted that Kaluza-Klein=like theories suffer severely from a 
"no-go" theorem by Witten\cite{witten} which essentially says that Kaluza-Klein theories 
cannot manifet  mass protected fermions in the $1+3$ dimensions. 
We want to mention here, however, our work 
on a toy model, in which a spinor in $1+5$-dimensional space manifests its 
masslessness in $d=1+3 $ space and yet chirally couples through a Kaluza-Klein charge, which is indeed 
the spin in $d-4$ space, to the corresponding Kaluza-Klein gauge field, if an appropriate 
boundary takes care of the properties of a spinor\cite{hn06}, overtaking accordingly 
the "no-go" theorem. 
But should these troubles be overcommed we better use a model with 
- since it  should of course have at least 4 dimensions - 6, 10, 14, 18, ...
dimensions. 

The works of one of us has sincee long had some success with 14 dimensions.

%\section{Acknowledgement } 

\end{document}